\documentclass[conference,a4paper]{IEEEtran}
\IEEEoverridecommandlockouts

\usepackage{cite}
\usepackage{amsmath,amssymb,amsfonts}
\usepackage{algorithmic}
\usepackage{makecell}
\usepackage{array,balance}  
\usepackage{lipsum}
\usepackage{graphicx}
\usepackage{subcaption}
\usepackage{multicol,multirow}
\usepackage{array}
\usepackage{graphicx}
\usepackage{color,soul}
\usepackage{xcolor}
\usepackage{nicematrix}
\usepackage{tcolorbox}
\usepackage{tabularx, boldline, multirow}
\usepackage[boxed,ruled,vlined]{algorithm2e}
\usepackage{enumerate}
\usepackage{textcomp, gensymb}
\usepackage{float}
\usepackage{subcaption}
\usepackage{graphicx}

\usepackage{amsmath}

\allowdisplaybreaks

\usepackage{geometry}
\title{Sparse OFDM Design for Interference and Ambiguity Mitigation in Multi-Static ISAC}

\author{Navid~Amani,~Priyanka~Maity,~Musa Furkan Keskin,~Henk Wymeersch\\
\IEEEauthorblockA{
Department of Electrical Engineering, Chalmers University of Technology, Gothenburg, Sweden
}
\thanks{This research has been carried out in WiTECH Centre (DISCOURSE project) financed by VINNOVA and partner companies.}
}

\begin{document}
\bstctlcite{IEEEexample:BSTcontrol}
\maketitle

\begin{abstract}
The sixth-generation (6G) wireless networks promises the integration of radar-like sensing capabilities into communication infrastructure. In this paper, we investigate a multi-static sensing framework where half-duplex base stations (BSs) are assigned as either transmitter or sensing receiver nodes. We propose a randomized sparse resource allocation scheme based on orthogonal frequency division multiplexing (OFDM) waveform design tailored for the multi-static scenario to simultaneously mitigate inter-BS interference (IBI) and sensing ambiguities. The waveform design also ensures robustness against inter-symbol interference (ISI) and inter-carrier interference (ICI) via a judicious choice of subcarrier spacing according to the deployment of BSs. The potential ambiguity caused by sparse signaling is addressed through controlled irregularity in both time and frequency domains, with a negligible noise floor elevation. Simulation results demonstrate the effectiveness and resilience of the proposed design in the presence of multiple targets and clutter.
\end{abstract}


%
\IEEEpeerreviewmaketitle

\section{Introduction}

{T}{he} sixth generation (6G) of wireless communication is set to extend the capabilities of existing cellular networks beyond ultra-low latency and high-throughput data connectivity, by enabling the seamless integration of sensing and communication (ISAC)~\cite{dong2022sensing}. 6G is expected to enable a distributed ISAC paradigm~\cite{strinati2025toward} for real-time situational awareness. Full-duplex base station (BS) operation, required for radar-like sensing~\cite{barneto2019full}, remains challenging and calls for architectural redesign~\cite{smida2023full}. A widely adopted approach involves half-duplex transceivers where radar functionality is integrated by partitioning BSs into transmitters (TX), which emit sensing waveforms, and receivers (RX), which handle radar signal processing~\cite{ liu2024cooperative, sakhnini2024distributed}. This enables a multi-static sensing scenario where multiple nodes could collaboratively transmit and receive sensing signals~\cite{li2023toward, behdad2024multi}.

\begin{figure} \centerline{\includegraphics[width=0.9\columnwidth, trim=5cm 0cm 4cm 0cm, clip]{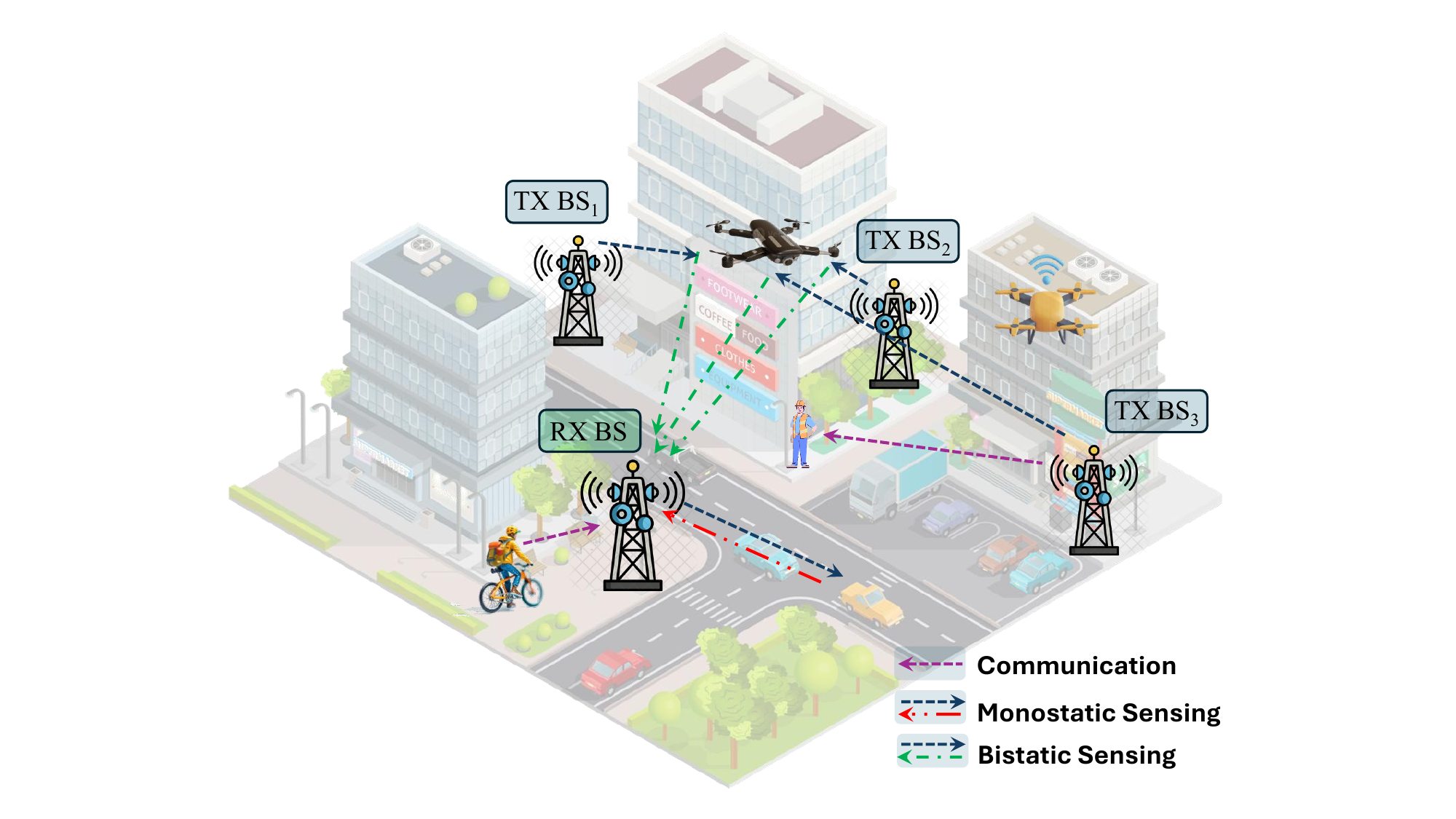}}
    \caption{Distributed multi-static sensing with BSs partitioned into transmitting and receiving nodes.}
    \label{fig:Dis_MIMO}
\end{figure}

In 5G standards, reference signals, such as positioning reference signals (PRS)~\cite{wei20225g}, are used for ranging of user equipments (UEs). The emergence of ISAC has triggered more advanced radar-like sensing capabilities, enabling the localization and tracking of both active entities (e.g., UEs) and passive objects (e.g., pedestrians, cyclists) with unprecedented accuracy~\cite{bourdoux20206g}. Conventionally, pilot signals are placed on equally spaced subcarriers, while data symbols occupy the intermediate subcarriers. Sparse sub-carrier selection is also proposed as a technique to reduce the analog-to-digital converter (ADC) sampling rate in OFDM sensing~\cite{lang2022ofdm}. This regular sparsity, however, results in range ambiguity and consequently limits the maximum unambiguous range for detection~\cite{staudinger2013sparse}. Similarly, ambiguity occurs in velocity estimation by regular sparse selection of OFDM symbols in time. Several studies have attempted to mitigate this issue by employing stepped pilot processing~\cite{ozkaptan2018ofdm} and joint exploitation of multiple reference signals~\cite{khosroshahi2025superposition, khosroshahi2024doppler}, by breaking the periodicity in either frequency or time domains. These methods lead to elevated sidelobes, forming a ridge-like noise structure across the range-velocity map (RVM)~\cite{hakobyan2019ofdm}. The authors in~\cite{mei2024coprime} introduced irregular pilot placement across both subcarriers and OFDM symbols to eliminate ambiguities in range and velocity domains. In \cite{mei2024coprime}, a monostatic sensing scenario with a single BS has been studied and it remains unclear how pilot allocation should be performed among multiple BSs in a distributed setup to circumvent both ambiguities and inter-BS interference. In addition, advanced signal processing techniques, such as compressive sensing, have been proposed to alleviate the ambiguity problem~\cite{rahman2020joint, wei2024multiple}. However, these methods often entail significant computational overhead.

In this paper, we propose a resource allocation scheme for multi-static ISAC systems that mitigates inter-BS interference and sensing ambiguity with negligible noise floor increase in RVMs. A portion of the OFDM grid is reserved for sensing, and multiple half-duplex BSs transmit dedicated sensing signals while one BS acts as the receiver. Orthogonal masking across BSs controls interference, and an irregular sparse allocation in both time and frequency mitigates sensing ambiguity by dispersing ghost targets across the RVM bins.

\section{System Model}
Consider a multi-static ISAC system consisting of $L$ TXs and $1$ RX, shown in Fig.~\ref{fig:Dis_MIMO}, all equipped with a single antenna. In the case of multiple RXs, the principle from this paper can be applied to each RX.  
\subsection{Transmit Signal Model}
\begin{figure} \centerline{\includegraphics[width=0.8\columnwidth, trim=9cm 3cm 10cm 4cm, clip]{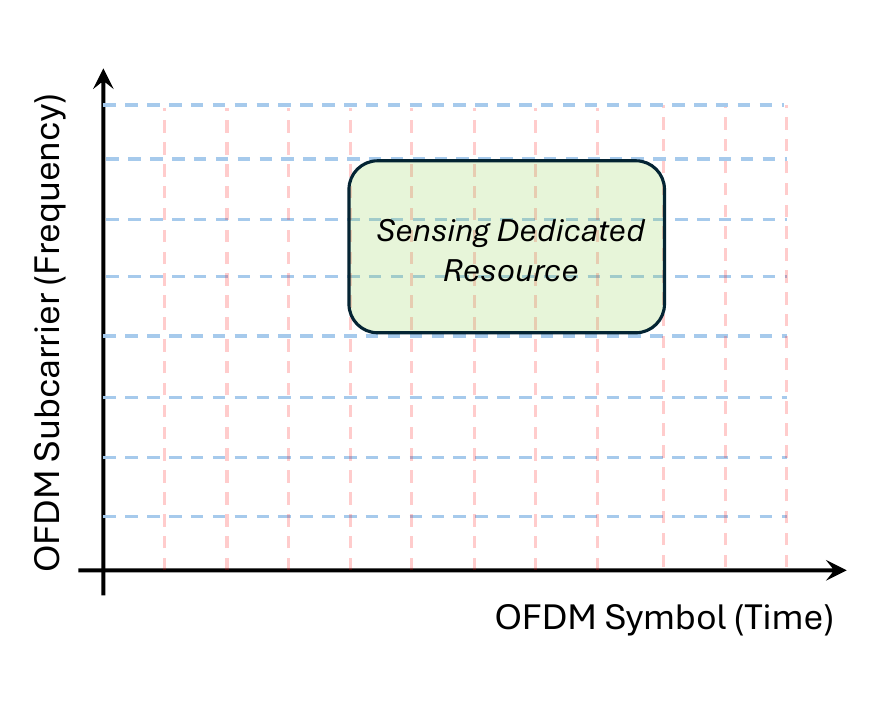}}
    \vspace{-4mm}\caption{Dedicated portion of the OFDM resource grid for sensing. Other resources are reserved for communication and not used for multi-static sensing. }
    \label{fig:Res_All}
\end{figure}

We analyze an OFDM frame consisting of 
$N_{\text{tot}}$ subcarriers and $M_{\text{tot}}$ symbols. In each OFDM frame, the transmitters allocate a predefined portion of the time-frequency resources dedicated to sensing, whereas the remaining resources carry communication data that are unknown at the receiver and hence excluded from sensing processing, as illustrated in Fig. \ref{fig:Res_All}. The complex baseband OFDM transmit signal from the ${{l}}$-th TX 
can be expressed as
\begin{align}
    s_l(t) = \sqrt{P_l}\sum_{m=0}^{M_{\text{tot}}-1} s_{m,l}(t),
\end{align}
where $P_l$ is the transmit power and 
\begin{equation}
s_{m,{{l}}}(t) = \frac{1}{\sqrt{N_{\text{tot}}}} \sum_{n=0}^{N_{\text{tot}}-1} x_{n,m,{{l}}} e^{j 2\pi n \Delta f t} g\left(\frac{t - mT_{\text{sym}}}{T_{\text{sym}}} \right),
\end{equation}
is the OFDM signal for the $m$th symbol, \( x_{n,m,l} \) represents the data or pilot on the \( n \)th subcarrier and the \( m \)th symbol, \( \Delta f = \frac{1}{T} \) is the subcarrier spacing with \( T \) denoting the elementary symbol duration, \( T_{\text{sym}} = T + T_{\text{cp}} \) is the total symbol duration including the cyclic-prefix (CP) time \( T_{\text{cp}} \), and \( g(t) \) is a rectangular pulse that takes the value 1 for \( t \in [0,1] \) and 0 otherwise. We set \( \mathbb{E} \left\{ |x_{n,m,{{l}}}|^2 \right\} = 1 \) and consider that sensing pilots are limited to a dedicated sensing resource of $N\le N_{\text{tot}}$ contiguous subcarriers and $M \le M_{\text{tot}}$ OFDM symbols, as shown in Fig.~\ref{fig:Res_All}. 
The upconverted transmit signal over the block of $M_{\text{tot}}$ symbols for $t \in \left[0, M_{\text{tot}} T_{\text{sym}} \right]$ can be written as \cite{keskin2025fundamental}
\begin{equation}\label{eq_passband_st}
\tilde{s}_l(t) = \Re \left\{  s_l(t) e^{j  2 \pi f_c t} \right\},
\end{equation}
where $f_c$ is the carrier frequency. 

\subsection{Channel Model}
The multiple TXs send data/pilot symbols to downlink users for communications while the RX collects the backscattered echoes for multistatic radar sensing. We assume the existence of $K$ targets, including mobile targets of interest and static clutter, in the environment. The baseband delay-Doppler channel for the $l$th TX-RX pair can be expressed as
\begin{equation} \label{eq_channel}
    h_l(\tau,\nu)= \sum_{k=1}^K \alpha_{l,k} \delta (\tau - \tau_{l,k})  \delta (\nu - \nu_{l,k}), 
\end{equation}
where $\alpha_{l,k} \in \mathbb{C}$, $\tau_{l,k}$ and $\nu_{l,k}$ represent the combined channel gain, bistatic delay and bistatic Doppler, respectively, of the two-segment path from the $l$th TX to the $k$th target and from the $k$th target to the RX. Here, $\alpha_{l,k}$ includes the effect of transmit power, radar cross section (RCS) and the overall path loss \cite[Eq.~(7)]{NF_Coh_Sens_2022}. For clutter, $\nu_{l,k}=0$.
 

\subsection{Necessary Conditions}\label{sec_conditions}
For the OFDM waveform, system parameters must be carefully chosen to prevent ISI and ICI. In particular, to ensure ISI-free communication, the CP duration must exceed the maximum delay spread in the channel. In the multi-static configuration, the multi-static delay spread can be formulated as:
$\tau_{\text{max}}=\max_{l,k} \tau_{l,k}$ and $\tau_{\text{min}}=\min_{l,k} \tau_{l,k}$, leading to the condition
$\tau_{\text{ms}} = \tau_{\text{max}} - \tau_{\text{min}} < T_{\text{cp}}$.
Furthermore, to ensure ICI-free operation, we consider the subcarrier spacing to be much greater than the maximum Doppler frequency, expressed as $\Delta f \geq 10 \max_{l,k} |\nu_{l,k}| $.
\subsection{Received Signal Model and Processing}
Based on the transmit signal in \eqref{eq_passband_st} and the sensing channel in \eqref{eq_channel}, 
the received signal at the RX for the $i$th sample of the $m$th symbol, after downconversion, CP removal and sampling at $t = mT_{\text{sym}} + T_{\text{cp}} +  iT / N_{\text{tot}}$ for $i = 0, \ldots, N_{\text{tot}}-1$, can be expressed as \cite{ICI_friend_foe_2021}
\begin{align}\label{eq_rec_bb2}
    y_{i,m} &= \sum_{l=1}^L \sum_{k=1}^K \alpha_{l,k} e^{-j (2 \pi f_c (\tau_{l,k} + \Delta t_l) +\Delta \phi_l)} e^{j 2 \pi m T_{\text{sym}} (\nu_{l,k} - f_{l}^{\text{off}})  } \nonumber  \\ &\times \frac{1}{\sqrt{N_{\text{tot}}}}  \sum_{n = 0}^{N_{\text{tot}}-1}  x_{n,m,l} \, e^{j 2 \pi  \frac{in}{N_{\text{tot}}}} e^{-j 2 \pi n \Delta f (\tau_{l,k} + \Delta t_l)},
\end{align}
where $\Delta t_l, f_{l}^{\text{off}}, \Delta \phi_l$ denote the time offset, carrier frequency offset and phase offset for the $l$th TX-RX pair, respectively. 
Let
\begin{align} \label{eq_steer_delay}
\mathbf{b}(\tau) 
&\triangleq  
\left[ 
1,\, e^{-j 2 \pi \Delta f \tau},\, \ldots, \right. \left. e^{-j 2 \pi (N_{\text{tot}}-1) \Delta f \tau} 
\right]^\mathsf{T} ,\\
\mathbf{c}(\nu) 
&\triangleq  
\left[ 
1,\, e^{j 2 \pi T_{\text{sym}}\nu },\, \ldots, \right. \left. e^{j 2 \pi (M_{\text{tot}}-1) T_{\text{sym}} \nu} 
\right]^\mathsf{T}~,\label{eq_steer_doppler}
\end{align}
represent the delay and Doppler steering vectors, respectively. An $N_{\text{tot}}$-point fast Fourier transform (FFT) is applied along the fast-time dimension (denoted by index $i$) of the received signal in \eqref{eq_rec_bb2} to switch from fast-time/slow-time to frequency/slow-time domain. 
Since only a subset of the total time–frequency resources is reserved for sensing, specifically $M$ OFDM symbols and $N$ subcarriers out of $M_{\text{tot}}$ and $N_{\text{tot}}$, respectively, we isolate the corresponding section of the received signal matrix for subsequent sensing operations. Aggregating these frequency/slow-time (i.e., symbol) domain observations into a matrix and incorporating the carrier-dependent phase term $e^{-j (2 \pi f_c (\tau_{l,k} + \Delta t_l) +\Delta \phi_l)}$ into complex channel gain $\alpha_{l,k} $, the resulting frequency/slow-time domain observation matrix, including additive noise, is obtained as
\begin{equation}
\mathbf{Y} = \sum\limits_{{{l}}=1}^{L}  (\mathbf{H}_{{l}} \odot \mathbf{X}_{{l}})  + \mathbf{\Omega} ,
\label{RXSignal}
\end{equation}
where
\begin{align} \label{eq_yy_i_rx_ap}
   \mathbf{H}_{{l}}  &= \sum\limits_{{{k}}=1}^{K} \alpha_{l,k} \mathbf{b}(\tau_{l,k}+ \Delta t_l ) \mathbf{c}^\mathsf{T}(\nu_{l,k}-f^{\text{off}}_l) \in \mathbb{C}^{N \times M},
\end{align}
is the channel matrix from the $l$-th TX to RX through the targets and clutter, where $\odot$ denotes the Hadamard (element-wise) product, $\mathbf{X}_{{l}} \in  \mathbb{C}^{N \times M}$ denotes the pilot symbol matrix corresponding to the subcarriers and OFDM symbols allocated to the $l$-th TX for sensing, and \( \mathbf{\Omega} \in \mathbb{C}^{N \times M} \) represents the noise matrix, the elements
of which are independently and identically distributed with
zero-mean and variance \( \sigma^2 = {N_0}({N \Delta f}) \), where \( N_0 \) is the noise power spectral density. To obtain the channel corresponding to the $l$-th TX, we apply reciprocal filtering \cite{keskin2025fundamental}, which estimates the channel by performing element-wise division of the received symbols by the transmitted symbols \cite{OFDM_CAF_SPM_2025}:
\begin{equation}
    \mathbf{\widehat{H}}_{{{l}}}= \mathbf{Y}\oslash \mathbf{X}_{{{l}}},
\end{equation}
where $\oslash$ represents element-wise division. In the case of a sparse OFDM waveform, where certain resource cells are not used for transmission, we set the channel estimate at those cells to $0$, i.e.,
\begin{equation} \widehat{H}_{n, m, {l}}= \begin{cases}\frac{{Y}_{n, m}}{x_{n, m, {{l}}}} & \text { if } x_{n, m, {{l}}}\neq 0, \\ 0 & \text { if } x_{n, m, {{l}}}=0,\end{cases}
\end{equation}
where ${Y}_{n, m}$ is the $(n,m)$th entry of the frequency-domain matrix $\mathbf{Y}$. Then, one can perform $N$-point inverse fast Fourier transform (IFFT) across the subcarriers (frequency) and $M$-point FFT across symbols (slow-time) to extract the RVM.
\section{Sparse Resource Allocation}
\label{sec:sensing-resource}
While the multi-static system can be free from ISI and ICI by imposing the conditions in Sec.~\ref{sec_conditions}, there are still two effects to consider: inter-BS interference (when $x_{n,m,l}\times x_{n,m,l'}\neq 0$ for some time-frequency resources and $l\neq l'$) and sidelobes of the ambiguity function (when $x_{n,m,l}=0$ for some $n,m$). Clearly, these two effects are in contrast, requiring careful and coordinated design of the pilots across the TX BSs.

\subsection{Irregular-Sparse OFDM Resource Allocation}

\begin{figure}
    \centering
    \subfloat[]{%
  \includegraphics[width=0.9\columnwidth, trim=7.5cm 6cm 5cm 4cm, clip]{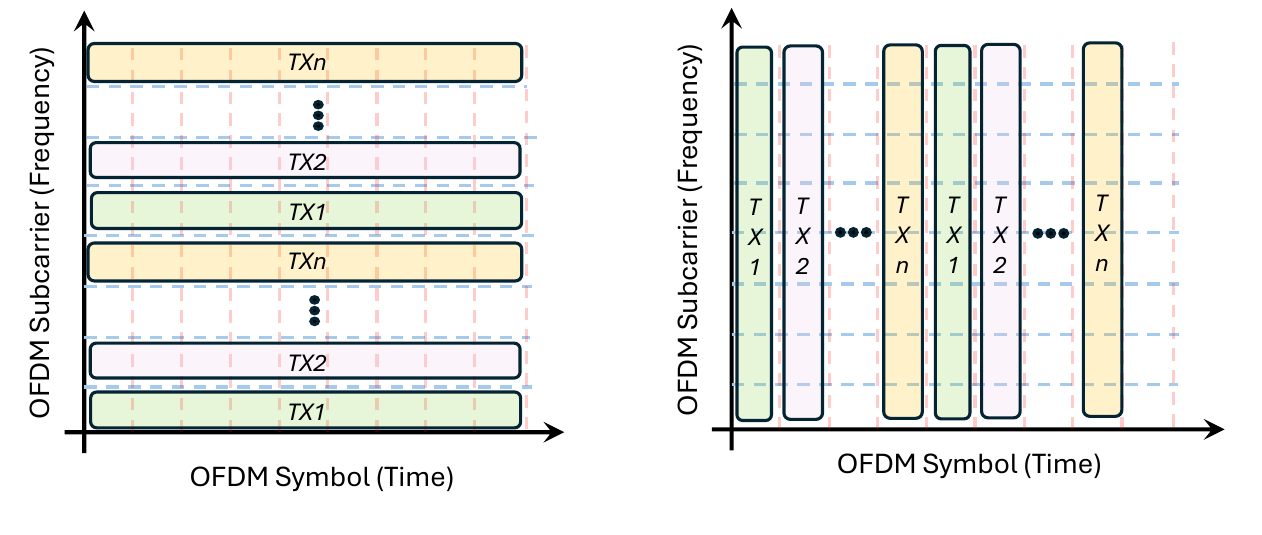}%
        \label{fig:orth_subfig_b}
    }
        \vspace{1em}
    \subfloat[]{%
  \includegraphics[width=0.9\columnwidth, trim=7.5cm 6cm 5cm 4cm, clip]{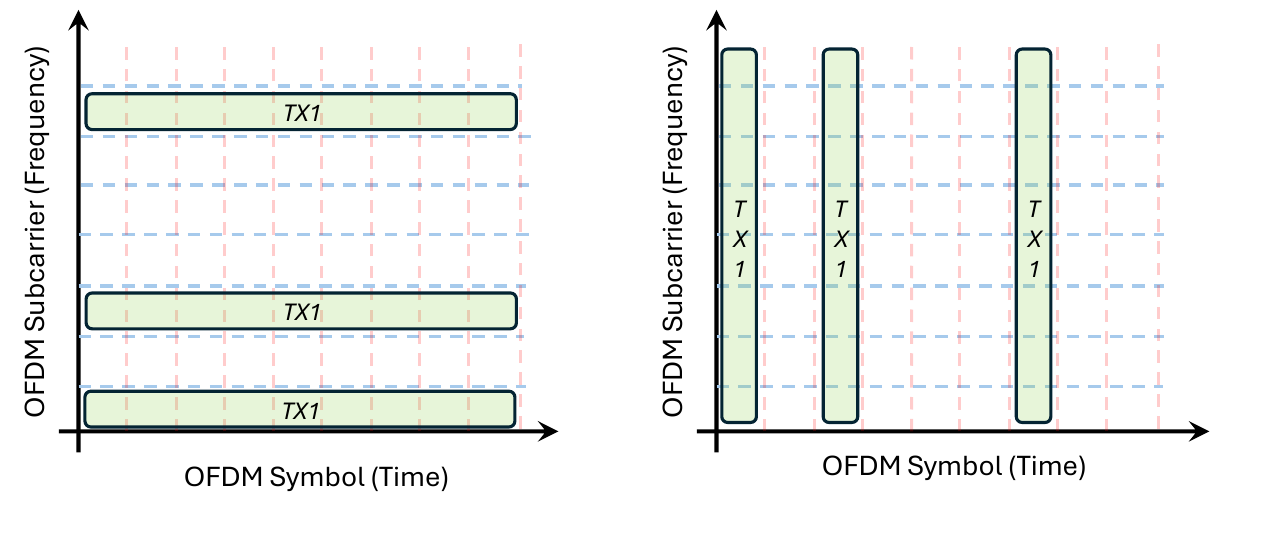}%
        \label{fig:orth_subfig_c}
    }
    \caption{Resource allocation scheme for managing inter-BS interference while preserving sensing resolution. (a) periodic allocation of resource segments per BS across the entire available resource window maintaining sensing resolution but introducing ambiguities, (b) aperiodic allocation of resource segments across the same window maintaining sensing resolution while suppressing ambiguities.}
    \label{fig:OrthogonalAllocation}
\end{figure}

In a multi-static scenario, to mitigate inter-BS interference and preserve sensing resolution, each BS can occupy small, non-contiguous portions of either time or frequency resources across the entire resource window. 
Time orthogonality, (see Fig.~\ref{fig:orth_subfig_b}, right), is achieved by assigning distinct transmission intervals, while frequency orthogonality, (see Fig.~\ref{fig:orth_subfig_b}, left), uses spaced subcarriers for each BS. Orthogonality patterns may be \textit{periodic} (Fig.~\ref{fig:orth_subfig_b}) or \textit{aperiodic} (Fig.~\ref{fig:orth_subfig_c}).  Periodic interleaving in frequency and time introduces range~\cite{khosroshahi2025superposition} and Doppler ambiguities~\cite{ khosroshahi2024doppler}, respectively.
The periodic sparsity is equivalent to multiplying the OFDM waveform (in frequency and/or time) by a windowed square wave with period $N_p$ and $N_a$ active resource cells. Since the RVM extraction applies the IFFT-FFT on the frequency-time grid, this multiplication corresponds to convolving their spectra in the range-velocity domain (i.e., RVM). When $N_a = N_p/2$, the square wave spectrum has sinc functions at odd harmonics; for $N_a \neq N_p/2$, both even and odd harmonics appear, producing ghost targets in the RVM at
\begin{equation}
    d_{\text{Ghost}} = d_{\text{Target}} \pm \gamma \times \Big ( \frac{d_{\max}}{N_p} \Big ),
\label{WindowedRect1}
\end{equation}
\begin{equation}
    v_{\text{Ghost}} = v_{\text{Target}} \pm \gamma \times \Big ( \frac{ 2 \times v_{\max}}{N_p} \Big ),
\label{WindowedRect2}
\end{equation}
where $d_{\max}={c}/{\Delta f}$ and $v_{\max} = {c}/{(4f_cT_{\text{sym}} )}$ are maximum unambiguous range and bistatic velcoity~\cite{de2023bistatic}, based on OFDM waveform parameters, and $\gamma\in \mathbb{Z}$ is the integer harmonic index, given that $ 0 < d_{\text{Ghost}} < d_{\max}$ and $\left| v_{\text{Ghost}} \right| \leq v_{\max}$.
While Fig.~\ref{fig:orth_subfig_b} illustrates periodic sparsity, the aperiodic approach in Fig.~\ref{fig:orth_subfig_c} reduces ambiguity by spreading ghost targets as side lobes in the RVM. Applying aperiodicity in only one domain, time or frequency, spreads ambiguities in the corresponding Doppler or range domain, producing an elevated ridge as sidelobe~\cite{hakobyan2019ofdm}, which may mask targets along that ridge. 
In summary, none of the allocations shown in Fig.~\ref{fig:OrthogonalAllocation} are suitable for multistatic sensing. 

\newcolumntype{M}[1]{>{\centering\arraybackslash}m{#1}}

\begin{table*}[!t]
\vspace*{0.1in}
\centering
\caption{Comparison of OFDM Resource Allocation Schemes}
\label{tab:resource_allocation}
\renewcommand{\arraystretch}{1.1}
\resizebox{0.95\textwidth}{!}{%
\begin{tabular}{|M{0.30\textwidth}|M{0.36\textwidth}|M{0.29\textwidth}|}
\hline
\textbf{Pilot Allocation} & \textbf{Advantage} & \textbf{Disadvantage} \\ \hline
Contiguous frequency or time allocation & No IBI & Reduction of sensing resolution \\ \hline
Frequency division with regular sparsity & No ambiguity in velocity, No IBI & Range ambiguity \\ \hline
Time division with regular sparsity & No ambiguity in range, No IBI  & Velocity ambiguity \\ \hline
Frequency division with irregular sparsity & No ambiguity in range and velocity, No IBI & Ridge in range profile \\ \hline
Time division with irregular sparsity & No ambiguity in range and velocity, No IBI & Ridge in velocity profile \\ \hline
Proposed 2D irregular sparsity & No ambiguity in range and velocity, no ridges, No IBI & Limited noise floor increase \\ \hline
\end{tabular}%
}
\end{table*}

\subsection{Randomized Sparse Resource Assignment}
\begin{figure}
    \centering

    \begin{minipage}[b]{0.45\linewidth}
        \centering
        \centerline{\includegraphics[width=\columnwidth, trim=11cm 5cm 12cm 5cm, clip]{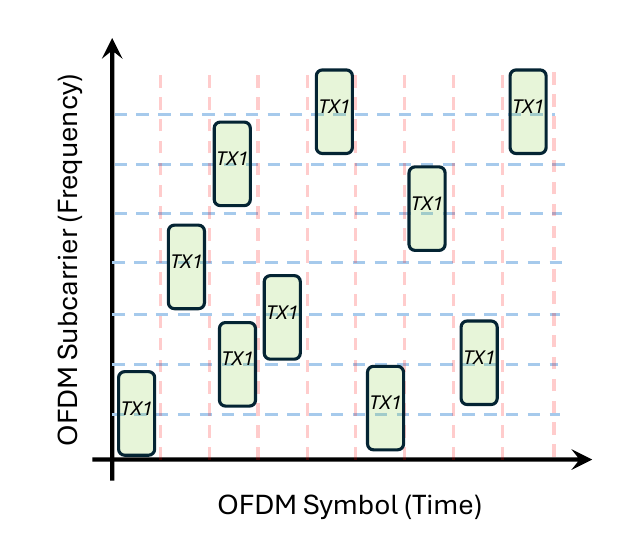}}
      
    \end{minipage}
    \hfill
    \begin{minipage}[b]{0.45\linewidth}
        \centering
        \centerline{\includegraphics[width=\columnwidth, trim=6cm 0cm 5cm 0cm, clip]{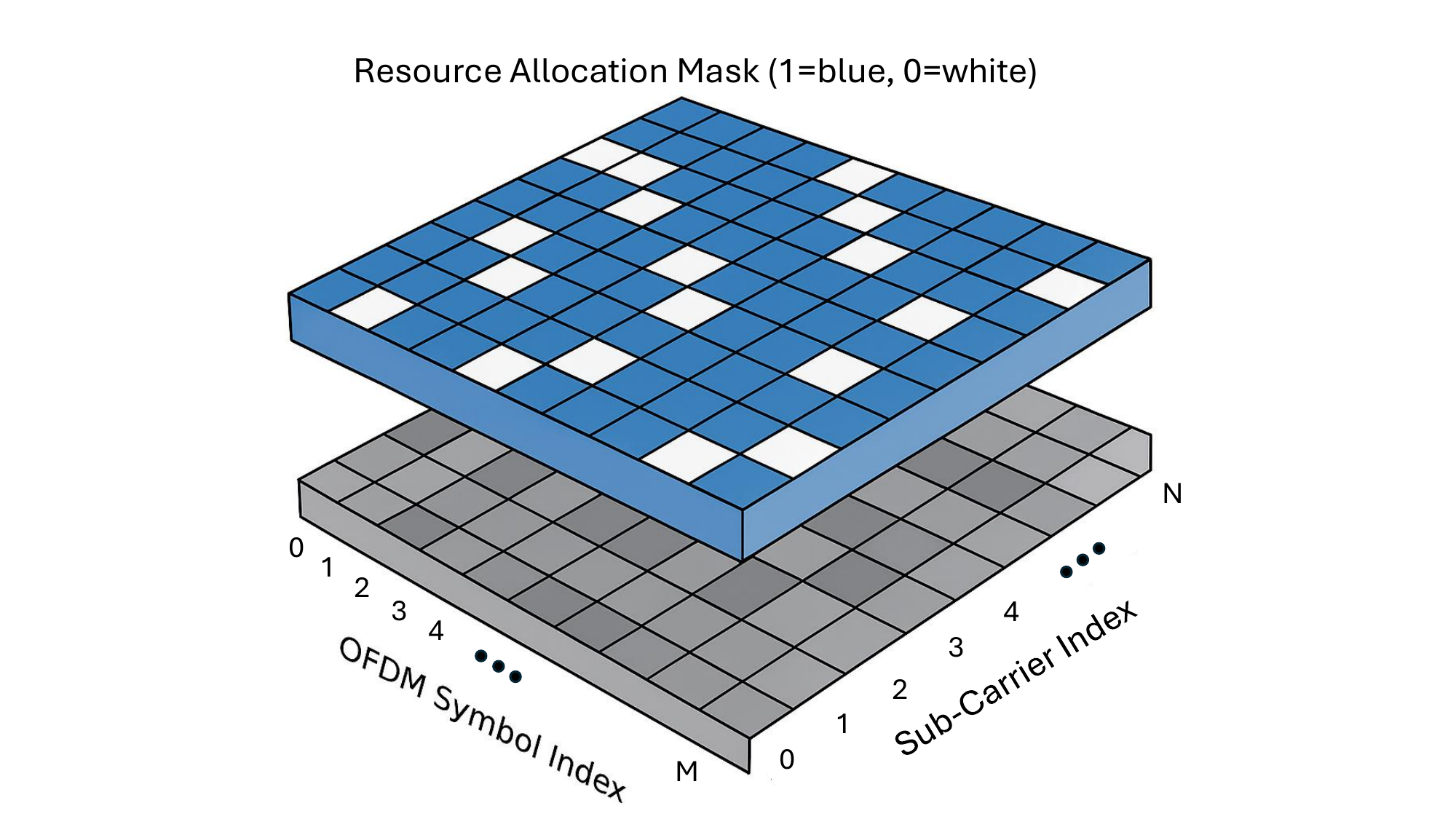}}

    \end{minipage}

    \caption{Sparse OFDM resource allocation, in both time and frequency domains, by assigning orthogonal masks to different transmitting BSs. }
    \label{fig:mask}
\end{figure}

To ensure aperiodicity in both subcarrier and symbol domains, we develop a randomized strategy. 
Consider $L$ transmitting nodes that share an OFDM grid with $N$ subcarriers
(indexed by $n\!\in\!\{1,\ldots,N\}$) and $M$ OFDM symbols
(indexed by $m\!\in\!\{1,\ldots,M\}$). 
For each time-frequency resource $(n,m)$, we independently draw a random
variable
$
Z_{n,m} \sim \mathrm{Unif}\{1,2,\ldots,L\},
$
and assign the resource to transmitter~$l$ if $Z_{n,m}=l$.
Let $A_l[n,m]\!\in\!\{0,1\}$ denote the resulting binary assignment mask, as in Fig.~\ref{fig:mask}:
\begin{align}
A_l[n,m] = 
\begin{cases}
1, & Z_{n,m}=l,\\[1mm]
0, & \text{otherwise.}
\end{cases}
\end{align}
By construction, every time-frequency resource is used by exactly one transmitter,
$\sum_{l=1}^{L} A_l[n,m] = 1$,
ensuring orthogonality across the transmitters.
The design has the following favorable property. 
Let $S_l=\sum_{n=1}^{N}\sum_{m=1}^{M} A_l[n,m]$ denote the number of
resources assigned to transmitter~$l$. It is easily verified that 
$\mathbb{E}[S_l] = {NM}/{L}$ and  $
\mathrm{Var}(S_l) = NM\,\frac{1}{L}\!\left(1-\frac{1}{L}\right)$. Hence, the relative standard deviation of $S_l$  decays as $\mathcal{O}(1/\!\sqrt{NM})$.   

The i.i.d.~assignment above maximizes aperiodicity in both time and frequency,
which is desirable for reducing structured range-Doppler ambiguities.
However, it does not guarantee that each subcarrier (row) or OFDM symbol
(column) contains exactly the same number of active transmitters.  
In general, we introduce a sparsity factor $\rho$, defined as the ratio of interleaved cells to the total number of resource cells in the OFDM grid:
\begin{align} \label{eq_rho}
    \rho = 1-\frac{1}{MNL} \sum_{l,m,n}A_l[n,m].
\end{align}
When $\rho=0$, all TXs fully use all the available OFDM resources, while $\rho=1-1/L$ corresponds to orthogonal usage.
Table~\ref{tab:resource_allocation} summarizes different resource allocation schemes along with their advantages and disadvantages. 

\subsection{Evaluating the Design}
\label{subsec:evaluate_design}

We evaluate the randomized sparse assignment through a CFAR-like
signal-to-interference-plus-noise ratio (SINR) measured on the per-TX
RVM.
Let $\widehat{\mathbf{H}}_l\!\in\!\mathbb{C}^{N\times M}$ be the
reciprocal-filter output for TX $l$ and let $\mathcal{F}_{N,M}$ denote the
IFFT (over subcarriers) and FFT (over OFDM symbols) used to form the RVM:
$
\mathrm{RVM}_l = \mathcal{F}_{N,M}\{\,\widehat{\mathbf{H}}_l\,\} \in \mathbb{C}^{N\times M}.
$
Select a \emph{cell under test} (CUT) at $(\hat\tau,\hat\nu)$ (a specific
range--velocity bin) associated with a known target for TX $l$. 
Define
\begin{align}
P_{\mathrm{sig}} \;=\; \sum_{(i,j)\in\mathcal{M}} 
\big|\mathrm{RVM}_l[i,j]\big|^2, 
\end{align}
where $\mathcal{M}$ is either the single CUT $\{(\hat\tau,\hat\nu)\}$ or a
small mainlobe set around it (to capture coherent energy). Let $\mathcal{G}$
be a guard set surrounding $\mathcal{M}$, and let $\mathcal{T}$ be the set of
$K$ training cells used to estimate the interference-plus-noise level, with
$\mathcal{M}\cap\mathcal{G}=\emptyset$, $\mathcal{M}\cap\mathcal{T}=\emptyset$,
$\mathcal{G}\cap\mathcal{T}=\emptyset$. The sample mean power from the training
cells is
\begin{align}
\widehat{P}_{\mathrm{I+N}} \;=\; 
\frac{1}{K}\sum_{(i,j)\in\mathcal{T}}\big|\mathrm{RVM}_l[i,j]\big|^2.
\end{align}
We then define
$
\mathrm{SINR} =
{P_{\mathrm{sig}}}/({\widehat{P}_{\mathrm{I+N}}})
$. 
When thermal noise is negligible, this metric reduces to a \emph{SIR}, where
$\widehat{P}_{\mathrm{I}}$ reflects the effective interference floor created by
cross-terms and ambiguity/ghost spreading due to multi-TX operation and sparse
masks. 

\section{Simulation Results}
\subsection{Scenario}
\begin{figure} \centerline{\includegraphics[width=0.85\columnwidth, trim=5cm 0cm 5cm 0cm, clip]{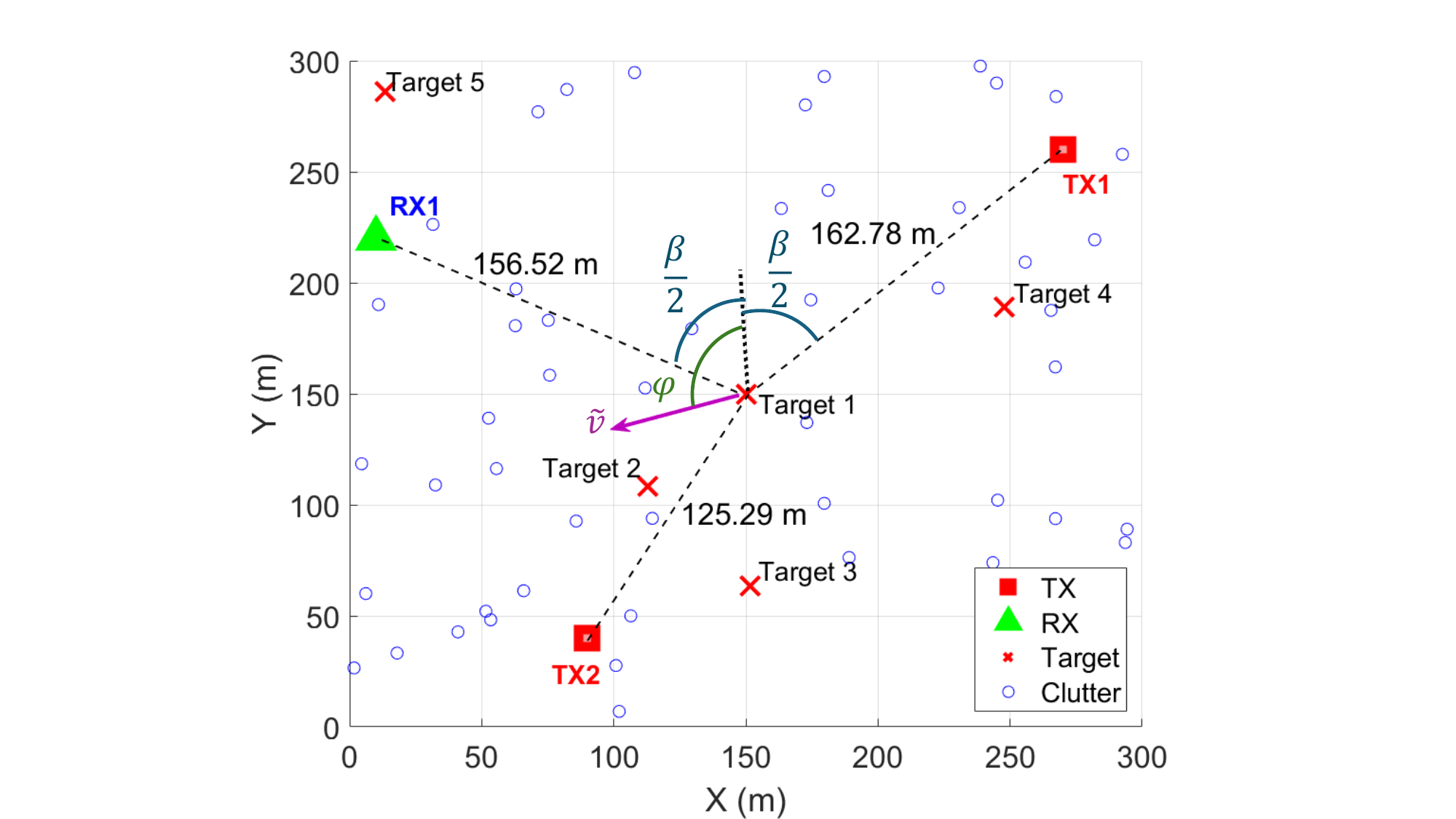}}
    \caption{Multi-static sensing scenario with two transmitting BSs and one receiving BS in the presence of multiple targets and clutter points.}
    \label{fig:Multi_Static}
\end{figure}

Within a square area of edge length $a=300$~m, BSs are randomly deployed with a minimum separation of $b=150$~m, as one realization is shown in Fig.~\ref{fig:Multi_Static} including multiple targets and clutter points. For simplicity, we assume perfect time and frequency synchronization between TX and RX BSs, i.e., $\Delta t_l=0$ and $f^{\text{off}}_l=0, \forall l$.
For $0 \leq b \leq a$, the maximum bistatic distance (BSD) can be achieved by: 
\begin{equation}
\text{BSD}_{\max} = a\sqrt{2} + \sqrt{
a^2 + (a-b)^2},
\end{equation}
while $\text{BSD}_{\min}=b$ occurs when the target appears on the base line connecting the TX and RX BSs, yielding $\tau_{\text{ms}}=2.03~\mu s$. To satisfy $T_{\text{cp}}>\tau_{\text{ms}}$ and given that $T_{\text{cp}}=0.07T_{\text{sym}}$, we have the condition $\Delta f \leq 34.4$ kHz.
For the maximum target velocity, we consider $50~\text{km/h} \sim 13.9~\text{m/s}$, representing moving objects such as bicycles or small vehicles. The Doppler frequency in the bi-static scenario with fixed TX and RX is given by~\cite{willis2005bistatic}: 
\begin{equation}
\nu = \frac{2v}{\lambda} = \frac{2\tilde{v}}{\lambda} \cos \phi~\cos\left(\frac{\beta}{2}\right),
\end{equation}
where $\phi$ is the angle of target motion relative to the bistatic bisector, $\beta$ is the bi-static angle, and $\tilde{v}$ is the target velocity (see Fig.~\ref{fig:Multi_Static}). 
So, the maximum Doppler frequency at $4.7~\text{GHz}$ is $435$ Hz. Hence, for ISI- and ICI-free operation:
\begin{equation}
4.35~\text{kHz} \leq \Delta f \leq 34.4~\text{kHz}.
\end{equation}
We select $\Delta f=30$ kHz, and therefore $T_{\text{cp}}=2.33~\mu s$.
For $100$~MHz bandwidth, we assume $273$ resource blocks and the total number of $273 \times 12=3276$ subcarriers. Within overall observation time of about $10~\text{ms}$, the number of OFDM symbols is 280. Table~\ref{tab:SimulPara} summarizes the OFDM parameters. 

\begin{table}
\centering
\caption{ OFDM Parameters}
\label{tab:SimulPara}

     


\begin{tabular}{ll}
\hlineB{3}
    Parameter & Value 
    \\ \hline \hline

         {Carrier frequency} & {$f_c = 4.7$ GHz} \\
         {Subcarrier spacing} & {$\Delta f = 30$ kHz} \\
        Total bandwidth & $\mathrm{BW}= 100$ MHz \\
         Number of subcarriers & ${N}= 3276$ \\    Symbol duration& $T=33.3\mu s$ \\
	 {CP} duration & $T_\mathrm{cp}=2.33\mu s$ \\
     OFDM symbol duration& $T_\mathrm{sym}=35.66\mu s$ \\
     Number of OFDM Symbols & ${M}= 280$ \\
     Frame Duration & $\mathrm{FD}= 10$ ms \\
     

	\hlineB{3}
\end{tabular}
\end{table}

\subsection{Sparsity-Based OFDM RVM}
\begin{figure}[htbp]
    \centering

    \begin{minipage}[b]{0.45\linewidth}
        \centering
        \includegraphics[width=\linewidth]{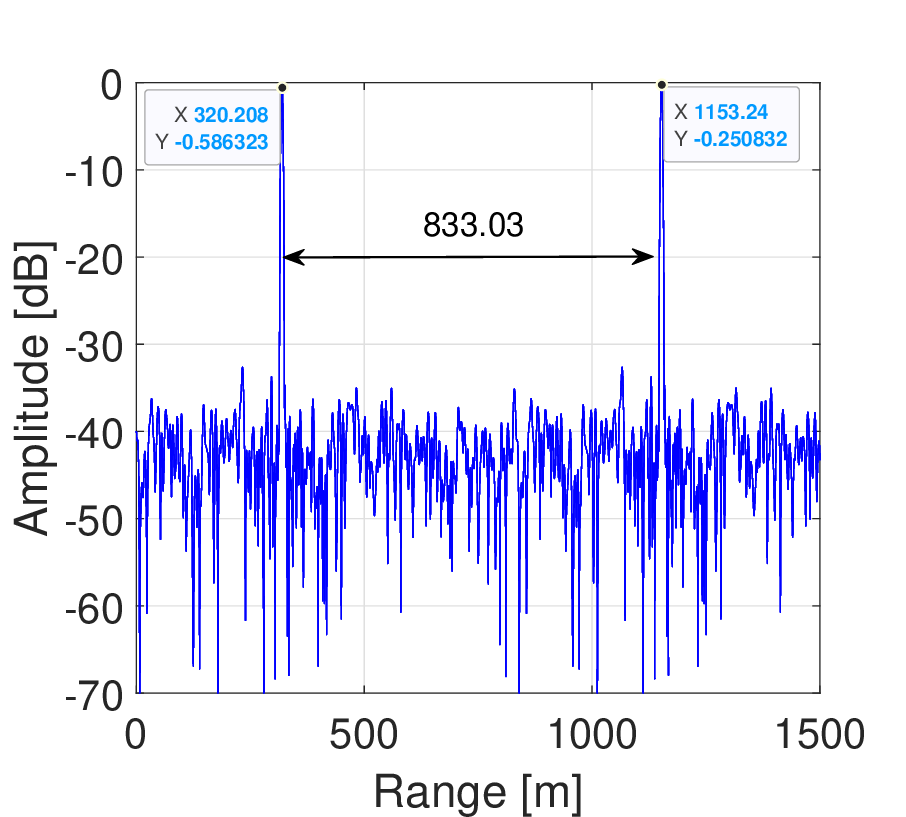}
        \par\vspace{1ex}
        (a)
    \end{minipage}
    \hfill
    \begin{minipage}[b]{0.45\linewidth}
        \centering
        \includegraphics[width=\linewidth]{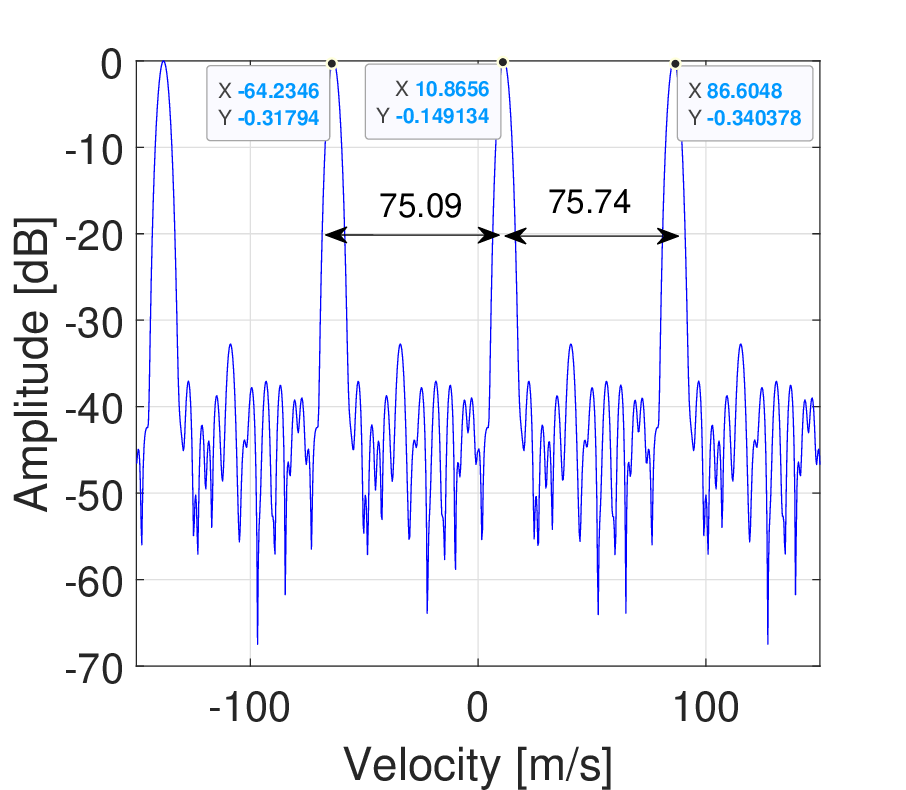}
        \par\vspace{1ex}
        (b)
    \end{minipage}

    \vspace{2ex}

    \begin{minipage}[b]{0.45\linewidth}
        \centering
        \includegraphics[width=\linewidth]{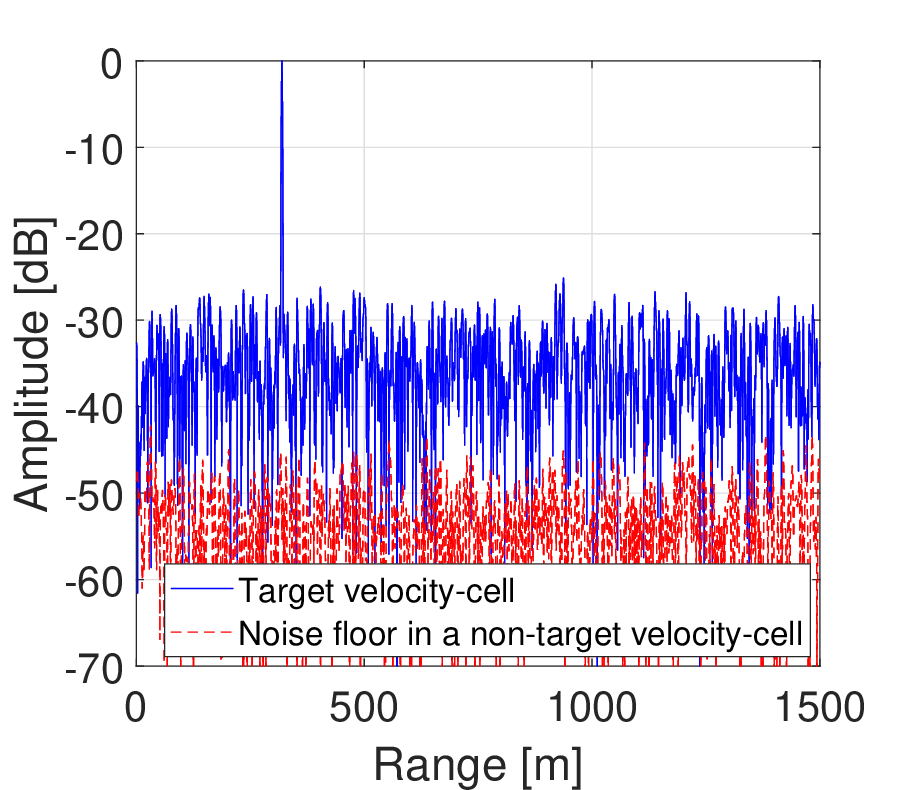}
        \par\vspace{1ex}
        (c)
    \end{minipage}
    \hfill
    \begin{minipage}[b]{0.45\linewidth}
        \centering
        \includegraphics[width=\linewidth]{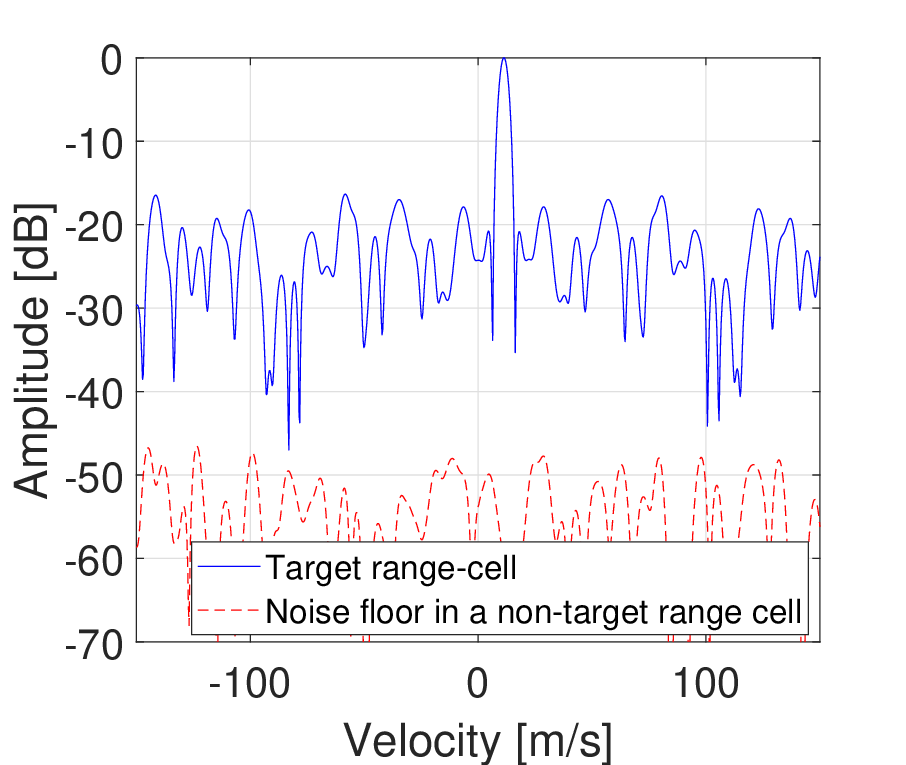}
        \par\vspace{1ex}
        (d)
    \end{minipage}

     \vspace{2ex}

    \begin{minipage}[b]{0.45\linewidth}
        \centering
        \includegraphics[width=\linewidth]{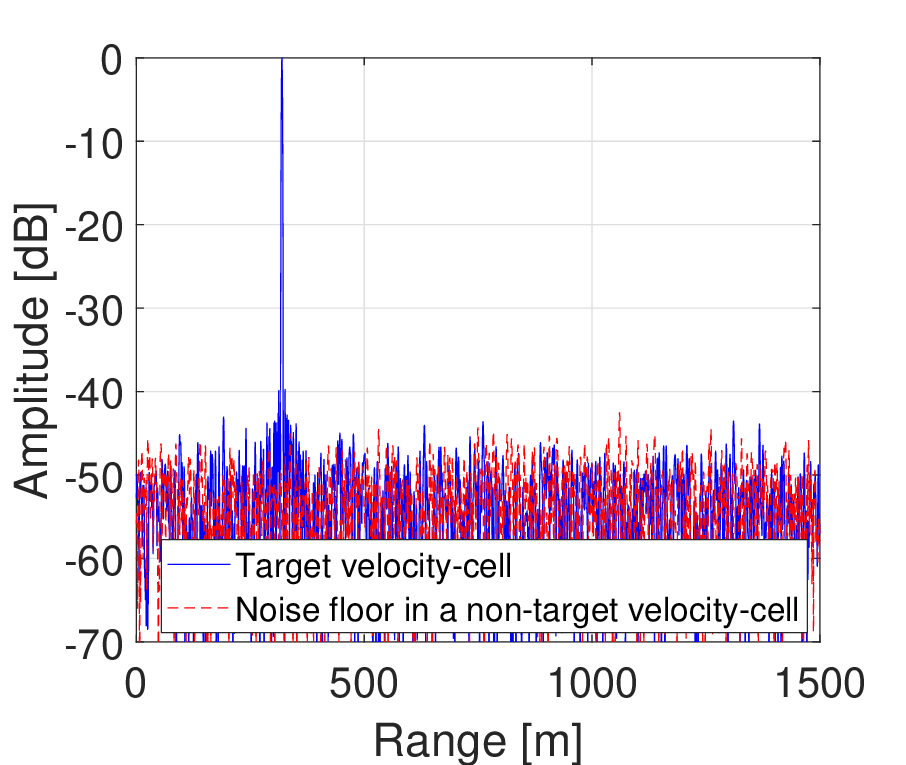}
        \par\vspace{1ex}
        (e)
    \end{minipage}
    \hfill
    \begin{minipage}[b]{0.45\linewidth}
        \centering
        \includegraphics[width=\linewidth]{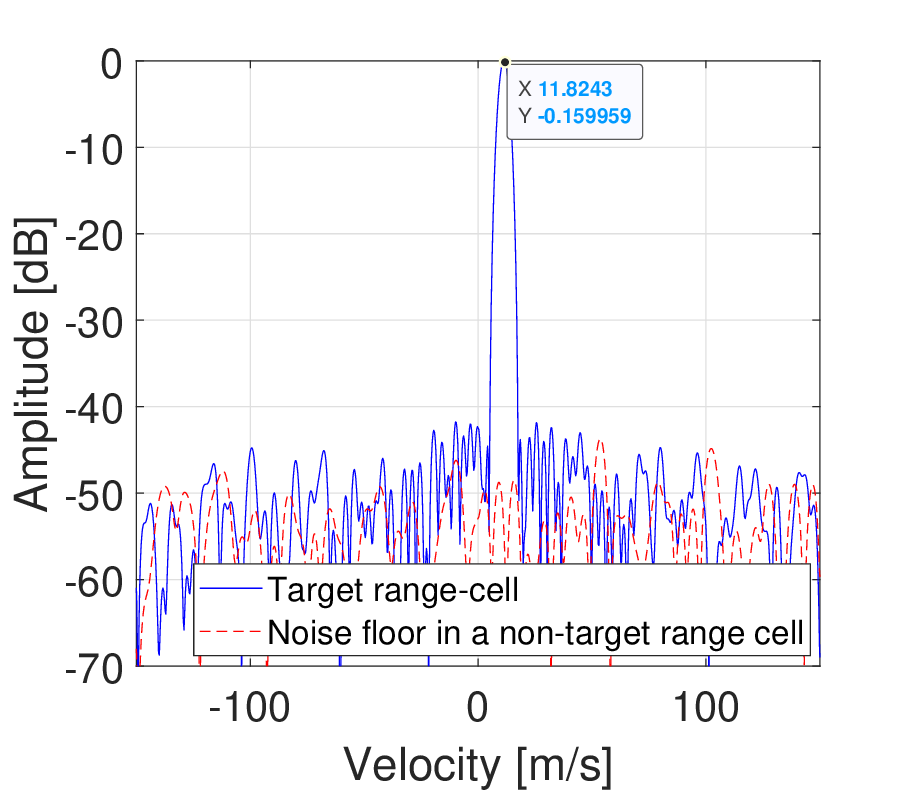}
        \par\vspace{1ex}
        (f)
    \end{minipage}

    \caption{(left) range-profile for the target velocity cell, (right) velocity-profile for the target range cell, Hamming window is applied. Periodic sparsity in the (a) frequency; (b) time, domain with $N_p=12$ equivalent to a PRS with a comb size equal to 12. Aperiodic sparsity in the (c) frequency; (d) time, domain. (e) and (f) aperiodic sparsity in both time and frequency domains with fully-orthogonal masks for two BSs.}
    \label{fig:RangeVelocityProfile}
\end{figure}

Fig.~\ref{fig:RangeVelocityProfile} illustrates the range and velocity profiles obtained from the TX1 signal for the scenario depicted in Fig.~\ref{fig:Multi_Static}, assuming that only Target~1 is present; all other targets and clutter are excluded. Target 1 has the bistatic distance of $319.3$ m and a velocity of $11$ m/s.
In the top row, due to periodic sparse resource allocation as shown in Fig.~\ref{fig:OrthogonalAllocation}a, with $N_p=12$ (equivalent to a PRS with comb 12), strong ghost targets appear. Periodic ghost targets with almost 833 m range spacing and 75 m/s velocity spacing, from the true target patameters due to ambiguity, follow equations~(\ref{WindowedRect1}) and~(\ref{WindowedRect2}), respectively. Aperiodic sparsity, in either frequency or time domain as Fig.~\ref{fig:OrthogonalAllocation}b, suppresses the high level ghost targets to a large extent; however, obvious noise floor increase is observed in the middle row. Ultimately, aperiodic sparsity in both frequency and time domains, as illustrated in Fig.~\ref{fig:mask}, suppresses the ambiguity issue with insignificant noise floor increase in the bottom row of Fig.~\ref{fig:RangeVelocityProfile}. 

\begin{figure*}[t]
    \centering
        \begin{subfigure}[b]{0.3\textwidth}
        \centering
        \includegraphics[width=\linewidth, trim=0cm 0cm 0cm 0cm, clip]{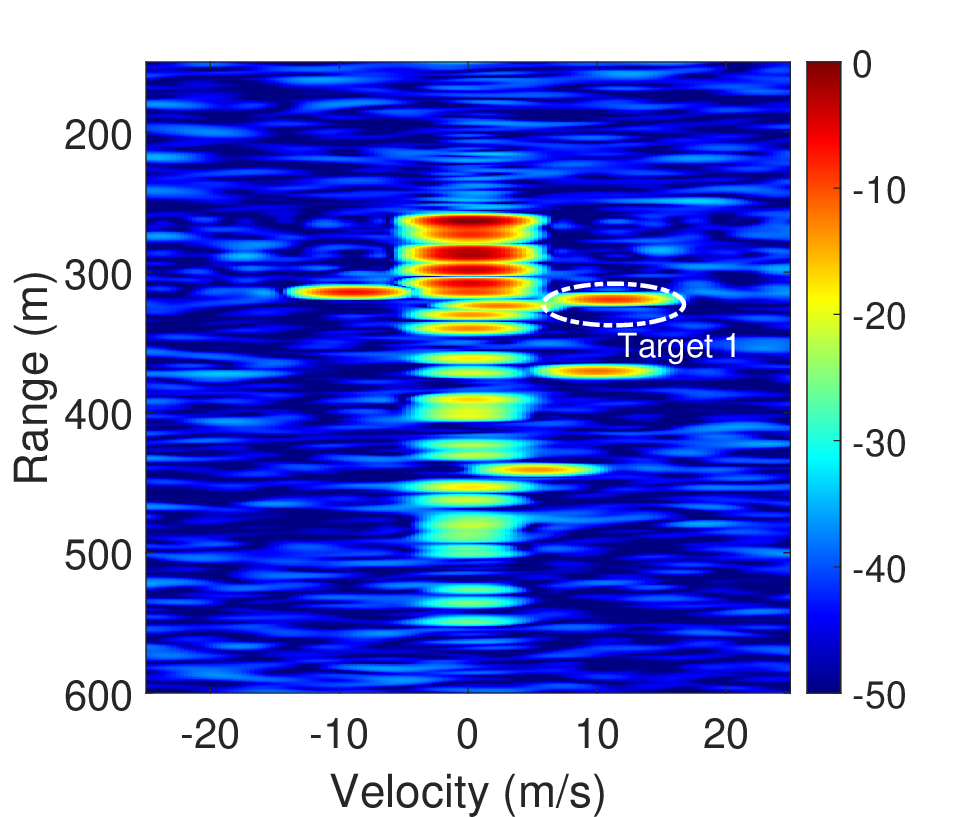}
        \caption{}
        \label{fig:sub_a}
    \end{subfigure}
    \hfill
    \begin{subfigure}[b]{0.3\textwidth}
        \centering
        \includegraphics[width=\linewidth, trim=0cm 0cm 0cm 0cm, clip]{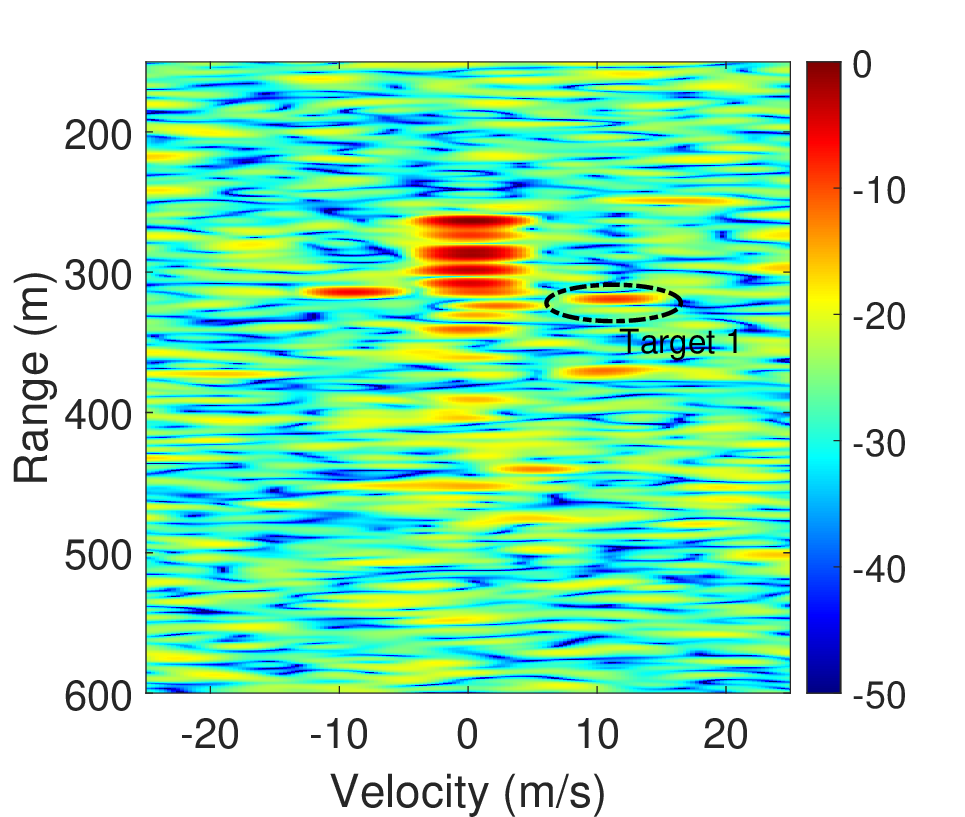}
        \caption{}
        \label{fig:sub_a}
    \end{subfigure}
    \hfill
        \begin{subfigure}[b]{0.3\textwidth}
        \centering
        \includegraphics[width=\linewidth]{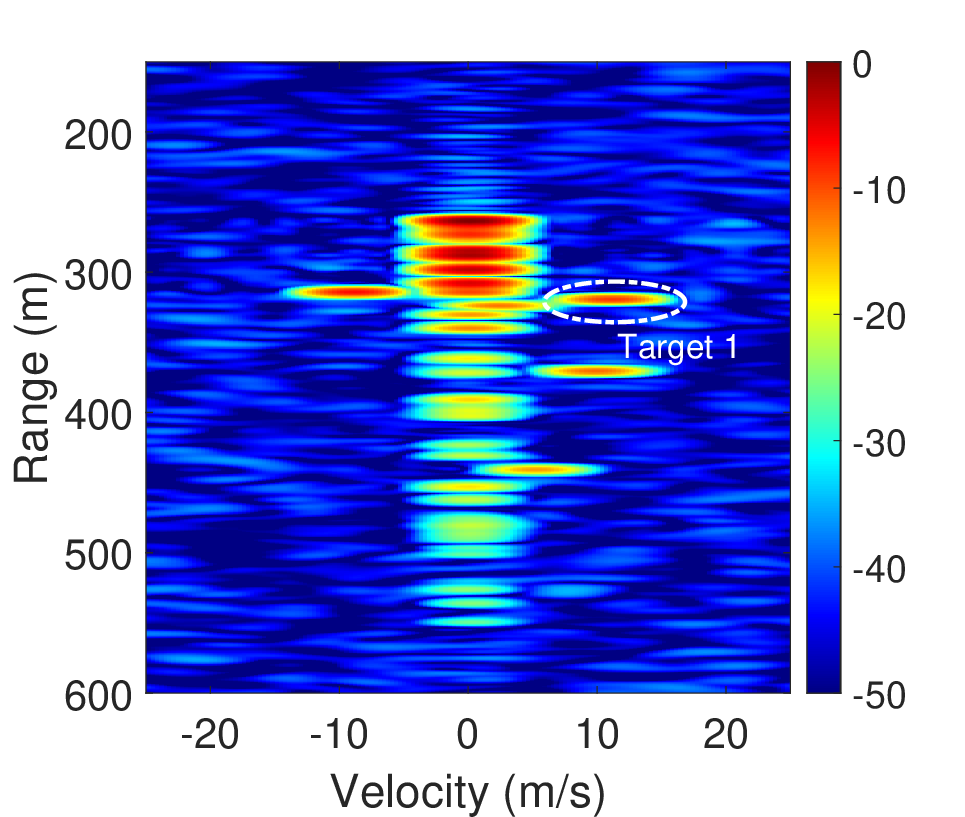}
        \caption{}
        \label{fig:sub_b}
    \end{subfigure}
    \caption{RVM with (a)  fully occupied OFDM resource cells by two TX BSs ($\rho=0$) with equal transmit power (0 dB power ratio), (b) fully occupied resource cells by two TX BSs ($\rho=0$) with 20 dB transmit power ratio, (c) orthogonal masks in OFDM resource allocation for two TX BSs ($\rho=0.5$) with 20 dB transmit power ratio.}
    \vspace{-6mm}\label{fig:RVMs}
\end{figure*}

We extend our simulation scenario by incorporating all 5 targets and 50 clutter points, and we extract the RVM from TX1 signal for both fully occupied OFDM resources by two TXs, in Fig.~\ref{fig:RVMs}a and b, and the proposed sparsity-based OFDM signaling in Fig.~\ref{fig:RVMs}c. Target and clutter points are modeled as isotropic scatterers. We also consider the case where BSs have different transmit powers, leading to problematic interference. Fig.~\ref{fig:RVMs}b illustrates that when TX2 transmits with 20 dB stronger power than TX1, the interference significantly contributes to the noise floor increase in the RVM associated with the TX1 signal. This interference is mitigated in Fig.~\ref{fig:RVMs}c by setting $\rho=0.5$ and random selection of resource cells in two orthogonal masks.



    \begin{figure}
    \centerline{\includegraphics[width=0.95\columnwidth]{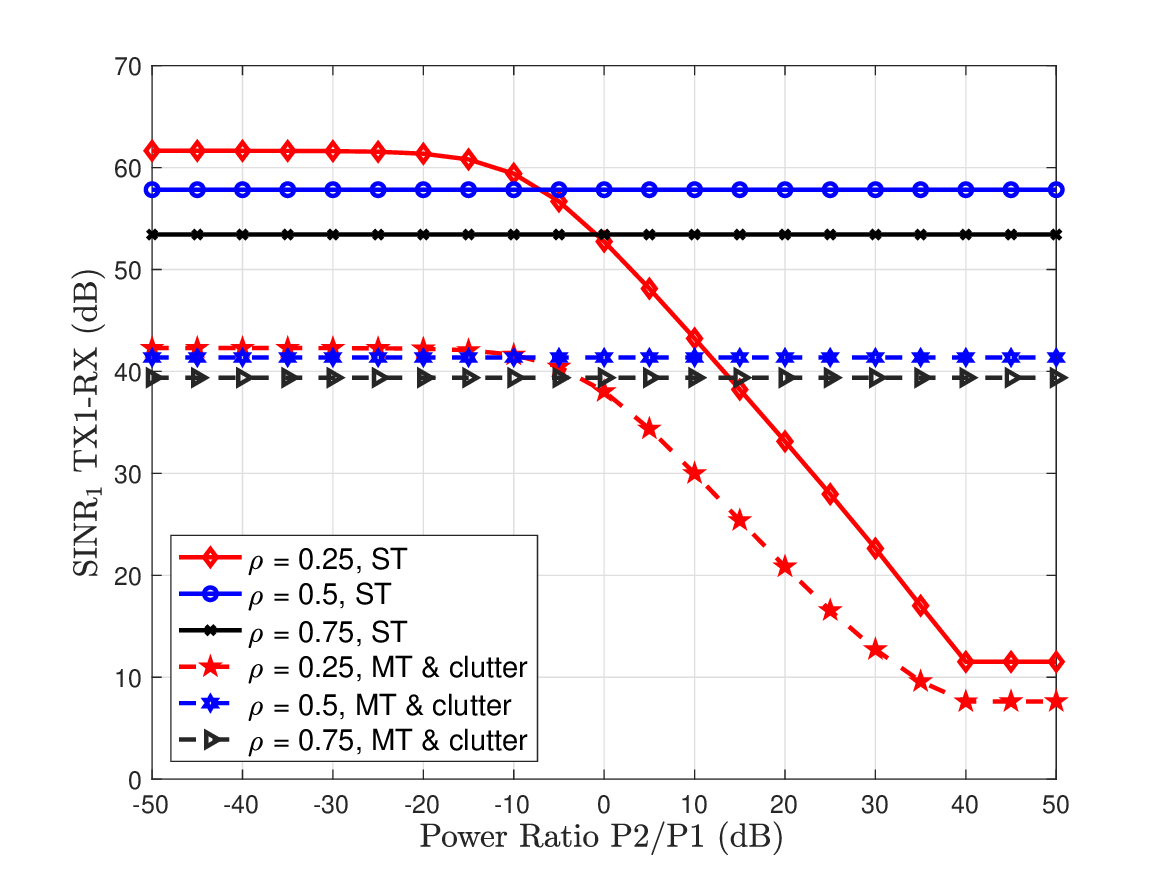}}
    \caption{SINR of a target under test in the RVM from TX1 signal while TX2 is considered as an interferer with varying transmit power. In the ST case there is only a single target in the scene, while in the MT \& clutter case there are multiple targets and clutter points randomly distributed in the scene.}
    \label{fig:SINR}
\end{figure}
Fig.~\ref{fig:SINR} illustrates the impact of the sparsity ratio $\rho$ on the SINR of Target~1 ($\text{SINR}_1$) from the TX1 signal, with the noise power set to zero, 
as a function of the transmit power ratio $P_2 / P_1$  between the TX1 and the TX2. We consider both the single-target (ST) and multi-target with clutter (MT \& clutter) cases. 
At power ratios below $0$ dB, TX1 dominates transmission power, and the overall $\text{SINR}_1$ at TX1 is primarily influenced by resource sparsity and ambiguity rather than interference since TX2’s contribution is negligible. In this regime, larger values of 
$\rho$ cause  ambiguity and degrade the $\text{SINR}_1$. As the power ratio increases, IBI becomes more significant, particularly for low $\rho$ allocation such as $\rho=0.25$ leading to a sharper $\text{SINR}_1$ decline. Sparsity factor $\rho=0.5$ represent orthogonal masks and achieves consistently better performance by minimizing IBI.

\section{Conclusion}

A sparsity-based OFDM resource allocation was proposed for multi-static sensing, where BSs are partitioned into transmitting and receiving nodes. Randomized interleaving of resource cells per BS across time and frequency mitigates inter-BS interference and ambiguities in RVMs. Simulations show minimal noise floor increase, making this approach a practical candidate for 6G ISAC use case. The proposed solution is critical for target detection on multiple RVMs corresponding to multiple transmitting BSs, enabling accurate localization and tracking of targets.
\balance

\bibliographystyle{IEEEtran}
\bibliography{refpaper1}

\end{document}